\documentclass[sigconf, screen]{acmart}

\usepackage{xcolor}
\usepackage{graphicx} 
\usepackage{subcaption}

\graphicspath{ {./images/} }
\AtBeginDocument{%
  \providecommand\BibTeX{{%
    \normalfont B\kern-0.5em{\scshape i\kern-0.25em b}\kern-0.8em\TeX}}}

\setcopyright{acmcopyright}
\copyrightyear{2018}
\acmYear{2018}
\acmDOI{10.1145/1122445.1122456}

\acmConference[Woodstock '18]{Woodstock '18: ACM Symposium on Neural
  Gaze Detection}{June 03--05, 2018}{Woodstock, NY}
\acmBooktitle{Woodstock '18: ACM Symposium on Neural Gaze Detection,
  June 03--05, 2018, Woodstock, NY}
\acmPrice{15.00}
\acmISBN{978-1-4503-XXXX-X/18/06}

\copyrightyear{2021}
\acmYear{2021}
\setcopyright{acmlicensed}\acmConference[FDG'21]{The 16th International Conference on the Foundations of Digital Games (FDG) 2021}{August 3--6, 2021}{Montreal, QC, Canada}
\acmBooktitle{The 16th International Conference on the Foundations of Digital Games (FDG) 2021 (FDG'21), August 3--6, 2021, Montreal, QC, Canada}
\acmPrice{15.00}
\acmDOI{10.1145/3472538.3472551}
\acmISBN{978-1-4503-8422-3/21/08}
\begin{document}


\title[Analyzing Players' Activity and Social Bonds in League of Legends During Covid-19 Lockdowns]{\#StayHome Playing LoL - Analyzing Players' Activity and Social Bonds in League of Legends During Covid-19 Lockdowns}


\author{Simone Petrosino}
\affiliation{%
  \institution{Institute of Interactive Systems and Data Science,}
  \institution{Graz University of Technology}
  \city{Graz}
  \country{Austria}}
\email{s.petrosino@tugraz.at}

\author{Enrica Loria}
\affiliation{%
  \institution{Institute of Interactive Systems and Data Science,}
  \institution{Graz University of Technology}
  \city{Graz}
  \country{Austria}}
\email{eloria@tugraz.at}

\author{Johanna Pirker}
\affiliation{%
  \institution{Institute of Interactive Systems and Data Science,}
  \institution{Graz University of Technology}
  \city{Graz}
  \country{Austria}}
\email{johanna.pirker@tugraz.at}

\renewcommand{\shortauthors}{Blinded, et al.}

\begin{abstract}
Humans are social beings. It is therefore not surprising that the social distancing and movement restrictions associated with the Covid-19 pandemic had severe consequences for the well-being of large sections of the population, leading to increased loneliness and related mental diseases. Many people found emotional shelter in online multiplayer games, which have already proven to be great social incubators. While the positive effect games had on individuals has become evident, we are still unaware of how and if games fostered the fundamental need for connectedness that people sought. In other words: how have the social bonds and interaction patterns of players changed with the advent of the pandemic? For this purpose, we analyzed one year of data from an online multiplayer game (League of Legends) to observe the impact of Covid-19 on player assiduity and sociality in three different geographical regions (i.e., Europe, North America, and South Korea). Our results show a strong relationship between the development of Covid-19 restrictions and player activity, together with more robust and recurrent social bonds, especially for players committed to the game. Additionally, players with reinforced social bonds---i.e., people played with similar teammates---were more likely to stay in the game even once the restrictions were lifted. 
\end{abstract}



\maketitle

\section{Introduction}
The Covid-19 pandemic has had a huge impact on people's lives and mental states. Social distancing caused a spiking feeling of loneliness~\cite{killgore2020loneliness}, correlated to depression and other serious health repercussions~\cite{killgore2020three}.
As a consequence, we experienced a drastic rise in social media usage. At the same time, entertainment platforms such as Twitch and Youtube also saw increasing viewership in terms of gaming activities~\cite{stephen_2020}. Apart from social media, games also saw a rapid activity increase. Steam, a game distributor, reported the highest peak of players in its history~\cite{King2020} and internet traffic related to gaming increased dramatically~\cite{lepido_rolander,shanley_2020}. People spent more time playing games to socialize and cope with stress~\cite{csener2021impact}. While social media overly exposed users to the dramatic situation world-wide, games represented a refuge and an escape from reality~\cite{coyle_allen_2020,Kriz2020}. Recent research on social play proving how multiplayer games are a social platform~\cite{Freeman2016MakingCommunity} fostering positive feelings~\cite{Pirker2018,Mandryk2020HowWellbeing}, such as connectedness~\cite{rogers2017motivational}, belonging~\cite{przybylski2010motivational}, and enjoyment~\cite{dabbish2012communication}. Those findings were also supported by recent research on how games impacted the live of people during Covid-19 and the social restrictions it brought. While researchers have already showed how player participation peaked in the past year~\cite{King2020} and how players drew psychological benefits in playing with others~\cite{johannes2020video}, little is known about how the social bonds created within games changed. In this context, we shifted from a player-centric analysis, focused on how players interact and benefit from the gaming experience, towards a network-centric investigation, by inquiring how Covid-19 restrictions impacted the social interaction patterns of players. To this end, we conducted an in-depth analysis on the League of Legends implicit social network, built from the in-game interactions of players during the Covid-19 crisis.

\subsection*{Research Questions and Contributions}~\label{research_question}
Covid-19 has had a significant impact on people's lives. This has had the result that individuals have relied on different ways of fulfilling their need for connectedness. In the light of previous findings that have presented arguments for games being great incubators of social relationships, which are also transferable offline, we expected an observable change in the player networks, built on the foundation of their in-game interactions in an online multiplayer game (i.e., League of Legends). In other words, this work aims at understanding how and whether online multiplayer have assisted people in difficult times. We pose the following research questions:

\begin{itemize}
    \item [RQ1.] How did player turnout and activity (\#matches and \#players) vary across the Covid-19 phases?
    \item [RQ2.] How did player social structures (clustering coefficient, communities, and focused player coefficient) vary across the Covid-19 phases?
    \item [RQ3.] Who are the players that approached the game during the Covid-19 lockdowns (churn rate, level of activity, and social involvement)?
    \item [RQ4.] How have player networks varied across different geographical regions?
\end{itemize}

By answering these research questions, we bring the following contributions to the Games User Research (GUR) and Human-Computer Interaction (HCI) fields. First, the analysis of variations in player engagement (RQ1) highlights how and whether players used online multiplayer games to cope with Covid-19 and lockdowns. The significant increase in player turnout and activity strongly suggests that multiplayer games occupy their newfound free time. This analysis confirms this trend for both new players and committed players taking refuge in LoL. While this rise in their activity signals that games assumed a significant role during lockdowns, this still does not connect with the social nature of games. Although games were an ideal means for safely filling time and respecting the government restrictions, we moved a step forward and analyzed how player social interaction patterns during social-distancing times. Understanding changes in the player social structure changes (RQ2) shows how lockdowns the social behaviors of players. One group of players, mostly comprised of committed LoL frequenters, made stronger and more recurrent bonds than thse prior to social distancing, and possibly finding in games the social connectedness that they lacked in life at that moment. Moreover, players starting to play during the lockdown weeks were more likely to remain in the game even in later stages if they were involved in stronger connections (RQ3). These findings not only support the conception of games being social platforms and of social interactions impacting retention but further connect online and real life. Events occurring in the real world have repercussions in the online game worlds, which modify their social norms and structures to fulfill their needs that reality cannot (temporarily) satisfy (e.g., connectedness). Finally, this study also compares the networks across different geographical areas (RQ4), providing evidence on how the different cultures impact the way players connect.

\section{Related Work}
In this section, we discuss evidence of games being great incubators of sociability and connectedness. Then we show how player telemetry data has been widely employed to gain knowledge of game status and player experiences. Finally, we show how players  have dealt with the pandemic through social gaming experiences.

\subsection{Games as social platforms}
Gaming can be a positive and healthy experience~\cite{Granic2014}. Playing, especially with other, can benefit different aspects of our life. It can help us to relax, reduce anxiety, and improve our mood~\cite{Ryan2006,Russoniello2009} and our cognitive abilities~\cite{Granic2014}. When it comes to sociability, it helps to learn positive skills related to cooperative and helpful behaviors, which are also practiced in the real life~\cite{Gentile2009,Granic2014}. The online gaming experience can produce a sense of connectedness and belongingness to a virtual community~\cite{Depping2017}. Thus, sociality represents a strong motivation for playing~\cite{Xu2011}, which can reduce the feeling of loneliness. As shown by other studies, solitude can lead to a broad spectrum of mental disorders~\cite{killgore2020loneliness,Stickley2016}. However, not all social games have the same positive effects on players. A feeling of belonging can only be achieved when social elements are well designed and aim at building an environment that connect players to one another~\cite{Ducheneaut2006,Depping2017}. As a result the focus of researcher interest is in analyzing player social interaction patterns to extract the rules and social norms, creating a prolific habitat for connectedness. Exploring sociability and behaviors of players can also be approached through the help of games' telemetry data.  Towards this, social network analysis comes in handy in investigating the social structure of players' social networks.

\subsection{Social Network Analysis in Games}
Social Network Analysis (SNA) is the discipline through which we can describe and analyze the relation between people~\cite{sage}. The literature on SNA and their application on games to learn about players and their social behaviors are wide. SNA is widely used to investigate many and different aspects of online games networks such as player interactions, player retention, the social structure of the game, engagement, influence, and communities~\cite{Pirker2018, Loria2020,Ducheneaut2007,Griffiths2011,Canossa2019}. Many studies focus on multiplayer games, which use a rich range of social features to build networks, such as friend lists or guilds~\cite{Canossa2019,Chen2008}. Others build the network starting from the match-making mechanism itself, linking two players if they played together~\cite{Losup2014, Mora-Cantallops2018, Jia2015}.
Studies on online multiplayer games provided evidence on the value of games as social outlets and how online relationships and reactions are often translated into the real world~\cite{Ducheneaut2006}. In this context,however we could not often verify whether events in the real world also impact virtual environments. The unprecedented pandemic has given us the chance to investigate how social dynamics evolved when people are forced to limit their physical, social relationships.

\subsection{Gaming during a pandemic}
An increase in the Covid-19 cases and deaths~\cite{who_stats} forced countries around the world to impose social distancing measures, which included closing borders, traveling limitations, bans on gatherings, and closing any meeting places or facilities~\cite{Chinazzi2020}. Governments generally made social contacts more difficult, even using mandatory stay-at-home orders (i.e., lockdown)~\cite{Alfano2020}. Many people faced the inconvenient situation of being more lonely and isolated than usual. Uncertainty about when the restrictions might end or become even more serious made mental health disorders and unhealthy behavior patterns severe problems~\cite{killgore2020loneliness,killgore2020three}. Since the beginning of the pandemic, the World Health Organization (WHO) has issued different suggestions about keeping ourselves mentally healthy~\cite{who_mental_health}. Among them, WHO supported online gaming as a tool for social distancing~\cite{canales_2020}. Remarkable is the case of \textit{Animal Crossing: New Horizons}, which went viral during the first pick of Covid-19 cases~\cite{khan_2020}. The game is a life simulation in which players are engaged in simple activities on their islands, such as growing plants, collecting fruits, or interacting with villagers. Players can have social interactions visiting other players' islands. Furthermore, a recent Oxford study shows how social games like \textit{Animal Crossing} have a positive effect on the mental health of players, reduce the feeling of loneliness, and foster well-being~\cite{johannes2020video}. The game encourages players to be positive and kind in their social behaviors, taking care of their islands and villages, helping other players, and giving gifts. Players can create their narrative to help them handle and express their emotions and feelings, which can be harder to express when dealing with social distancing~\cite{coyle_allen_2020}. Different to this are the cases of players having weddings, social events, and even funerals~\cite{kobek_patrick}. Mental health problems are also caused by unhealthy behaviors linked to a sedentary existence. While exergames--i.e., integrating physical activity and gaming features--are often recommended to promote a healthy and sporty lifestyle, they can also boost our social experiences when combined with social games. Through social exergames, players could fight anxiety and sedentary habits while sharing their results to connect with and keep in contact with other people~\cite{viana2020exergames}. If exergames promote a healthy lifestyle and connection while maintaining a physical distance, this was more difficult for location-based games (e.g., Pokemon GO). Location-based games rely on the real players' positions to offer different game contents. Due to the pandemic, they got special updates to keep on offering players an engaging social experience but avoiding them to have external social activities~\cite{laato2020did}. Games like Animal Crossing and exergames offer mechanics close to real-world interaction, answering to that desire for social interaction, such as having a conversation, hanging out with friends, or having physical actives. We know less about more ``classic'' games that do not entirely rely on those mechanics. We contribute to understanding the relationship between games and reality by filling this knowledge gap and investigating the effect of forced social distancing on players' activities and participation in a multiplayer online game (League of Legends).

\begin{table}[]
\centering
\caption{Countries in each League of Legends area}
\label{table:server_stats}
\begin{tabular}{lll}
\hline
Area   & \textbf{\#Countries} & \textit{\textbf{Countries}}                                                                                                                \\ \hline
\textit{EUW} & 11                  & \begin{tabular}[t]{@{}l@{}}Portugal, Spain, Germany, \\ Italy, France, Belgium, Netherland, \\ UK, Ireland, Austria, Switzerland\end{tabular} \\
\textit{NA}  & 2                   & Canada, U.S.                                                                                                                               \\
\textit{KR}   & 1                   & South Korea                                                                                                                                \\ \hline
\end{tabular}%
\end{table}

\section{Materials and Methods}
In the following section, we detail the process through which we gathered and analyzed player data. Our observation period extends from September 2019 to September 2020. In this time frame, the pandemic went through different phases. At the same time, countries reacted in different ways based on the virus spread. To better understand players' gaming behaviors throughout the year, we first need to take a closer look at restrictions and social distancing measurements, which differ in every country. Moreover, later in this section, we explain why we focused on three specific geographical areas (Table~\ref{table:server_stats}) and how we built the players' networks.

\subsection{Covid-19 Lockdown and Restrictions}
~\label{subsec:covid_phases}
During our observation period, each country has assumed specific strategies. We identify three phases, based on government reports and Covid-19 official spreading data that we detail later for each region:
\begin{itemize}
    \item \textit{Pre-Lockdown}. This phase includes the weeks before any form of lockdown or restrictions. In this period, the first news about Covid-19 appeared in the media, as the cases increased daily.
    \item \textit{Lockdown}. This phase includes the weeks in which restrictions were issued. For instance, movements within the country or cities of residence, and, in some areas, the limitations culminated into full-fledged lockdowns. During this period, gatherings were discouraged or prohibited, and shops, restaurants, and entertainment facilities were closed. Please note that we call this phase \textit{lockdown}, but it also indicates a phase in which restrictions were applied although a real lockdown did not exist (i.e., in South Korea).
    \item \textit{Post-Lockdown}. This phase does not necessarily overlap with a substantial decrease in Covid-19 cases but is primarily related to a significant decrease or even completion of lockdowns and movement restrictions. 
\end{itemize}

The more precise temporal collocation of each phase changes with the geographical area, as it is based on how countries manage the virus spread.

\subsubsection*{West Europe (EUW)}
In western Europe, the first declared Covid-19 case appeared by the end of January~\cite{timeline_europe}. Nevertheless, it was not until early March that European countries began to activate serious lockdown measures. 
The first country to enter into a lockdown was Italy~\cite{italy_lockdown}, but within a few weeks, over 250 million people were in lockdown~\cite{europe_lockdown}. 
Social distancing measures varied slightly from country to country. All the measures, however, included social distancing, assembly bans, border controls (inside and outside the country), house arrest regulations, closure of nonessential stores, museums, entertainment venues, and schools. Lockdowns and other forms of restrictions were lifted slightly across Europe starting in early May due to the decrease of cases and deaths~\cite{lift_europe}. We divided our observation period in three phases as follows as follows:
\begin{itemize} 
    \item Pre-Lockdown: week 1 - 27  (September 4, 2019 - March 11, 2020). This phase ranges from weeks where Covid-19 was not yet known to the first cases detected across Europe~\cite{timeline_europe}.
    \item Lockdown: week 28 - 37  (March 11, 2020 - May 20, 2020). From week 28, most of the Europe West countries entered the lockdown/restrictions phase~\cite{covid_start_timeline}.
    \item Post-Lockdown: week 38 - 52 (May 20, 2020 - September 2, 2020). From week 38, most Western European countries lifted the heavy restrictions and hard lockdown~\cite{lift_europe}.
\end{itemize}

\subsubsection*{North America (NA)}
For countries in North America, the breakdown of phases was slightly different. As in Europe, the first cases were confirmed both in Canada and the U.S. by the end of January. Social distancing measures in America were less strict than in Europe. They included social distancing, bans on gatherings, border controls (within and outside the country), closure of nonessential shops, museums, entertainment venues, and schools. Following those restrictions, masks and other protective devices became mandatory in most of North America~\cite{mask_usa,mask_canada}. A state of emergency was declared in all Canadian provinces by the end of March. The same happened for the 50 U.S. states that declared the state of emergency a few days after the U.S. president's emergency declaration~\cite{the_white_house}. However, in the U.S., the first stay-at-home orders did not start until late March~\cite{usa_start}. Both Canada \& U.S. started to lift the restrictions in mid-May~\cite{end_canada,machine_end_usa}. In Canada, this is related to a decrease in cases and deaths. Taking this information into consideration we divided and modified the three phases in the following way:
\begin{itemize}
    \item Pre-Lockdown: week 1 - 28 (September 4, 2019 - March 18, 2020). This phase goes from the weeks where Covid-19 was still unknown until the state of emergency was declared both U.S. \& Canada~\cite{covid_start_timeline}.
    \item Lockdown: week 29 - 39  (March 18, 2020 - June 3, 2020). From week 29, all Canadian provinces and most U.S. states declared a state of emergency and started to impose restrictions~\cite{covid_start_timeline}.
    \item Post-Lockdown: week 40 - 52  (June 3, 2020 - September 2, 2020). From week 40, all Canadian provinces lifted most of the restrictions, 23 U.S. states lifted the stay-at-home order, and most U.S. States reopened nonessential shops and entertainment places~\cite{machine_end_usa,machine_end_canada}.
\end{itemize}

\subsubsection*{South Korea (KR)}
South Korea fought Covid-19 in a different way. It is the area with least restrictions and with no imposed lockdown. A first Covid-19 patient was reported on January 20, 2020~\cite{first_case_korea}. Less than one month after, the ``Infection Disease Alert'' was raised from level 0 to 4~\cite{raised_alert_korea}. Compared to European and North American countries, South Korea did not rely on enforced lockdowns, curfews, or similar restrictions. Instead, reliance was placed on a massive campaign of testing and on an advanced contact tracing process to trace people who might be infected\cite{contact_tracing_korea,contact_tracing_korea2}. In addition, the population was advised to distance themselves socially, schools and entertainment venues were closed the moment the pandemic broke out. In addition, South Korea immediately closed its borders to many other countries~\cite{ministry_immigration_korea}. South Korea successfully contained the first Covid-19 wave thanks to its previous experience with a similar disease, the MERS, outbreak of 2015~\cite{mers_korea}. Due to the lack of strict and clear restrictions, which could help to separate a "before" and "after," we decided to divide South Korea into only two phases:

\begin{itemize}
    \item Pre-Lockdown: week 1 - 24  (September 4, 2019 - February 19, 2020). This phase goes from the weeks in which Covid-19 was still unknown to when the ``Infection Disease Alert'' was raised.
    \item Lockdown: week 25 - 52  (February 19, 2020 - September 2, 2020). From week 25, the alert was increased, the country's border control increased, and the school and other facilities closed~\cite{timeline_korea}
\end{itemize}

\subsection{League of Legends (LoL) - Data and Preprocessing}
League of Legends (LoL) is a multiplayer online battle arena (MOBA) developed and supported by Riot since 2009. In every match, two teams compete with each other intending to destroy the main base of the opponent team (Nexus). The teams are composed of five members, and each member has a specific role defined before the matchmaking phase allowing the algorithm to create a balanced team. Teams fight on 3D maps, divided into three lanes (Top, Mid, Bottom) and a jungle where resources can be harvested by killing monsters. Players can communicate through an internal chat or a ping system based on emojis, animations, or map marks. 

Players are ranked according to their skills in seven tiers: Iron, Bronze, Silver, Gold, Platinum, Diamond, and Master. Each tier (except Master) has four subcategories named as divisions. These can play in ranked and unranked matches, where unranked matches are mostly used for training purposes or special in-game events. The matchmaking for competitive matches is based on the rank/division system. The higher the tier, the smaller the choice of players with a different rank to play with. For example, players in the Bronze tier can play with both Bronze and Silver tier players, whereas Platinum players can only play with Gold or Platinum tiers. 

Besides being divided by rank and expertise, LoL players are also associated with one of eleven geographical areas. When they register, players need to specify the geographical area of reference, which represents the server location they will connect to. Players in different areas cannot play together. Eventually, a player can transfer the own account to another server paying for it. Creating a new account in a different area is always possible and is free.

\subsubsection*{Data Collection \& the Players' Network}
To perform the study, we collected data for : (a) players' activities, in terms of how many matches they played, and (b) their in-game connections describing with whom they played. In the following, we first explain how we chose the LoL players included in the dataset and how we built the (implicit) player network. We obtained the material used for this study from the Riot Official API \footnote{https://developer.riotgames.com/apis} system.

Players can participate in a match only with players registered to the same area---i.e. the geographical position of the server. We focused our analysis on three areas, North America (NA), Western Europe (EUW), and South Korea (KR). Table \ref{table:server_stats} shows the countries included in the server. Two reasons were the motivation behind our decision. First, at the data collection date (September 10, 2020), those servers represented the biggest LoL communities~\cite{websitelol}. Second, these three areas represent very well the different restrictions approaches that countries worldwide have taken to fight Covid-19. 

We modeled the LoL player network as an undirected graph. Nodes represented players connected by an edge between if they played in a match as teammates. This way of modeling players' in-game implicit relationships is common in the GUR literature, in studies analyzing LoL~\cite{Mora-Cantallops2018} and other team-based games~\cite{Loria2020,Canossa2019}. The gathering processes we followed, were identically repeated for each geographical region (i.e., EUW, NA, and KR).

Besides choosing the geographical area, we also focused on players belonging to the Bronze, Silver, and Gold tiers. Hence, we excluded the lowest ranking (Iron) and the top rankings (Platinum, Diamond, and Masters). The choice was motivated by these three tiers representing the majority of the player population\cite{Mora-Cantallops2018}. Additionally, players in these tiers were the ones most likely to play together due to the LoL matchmaking system. We thus limited the sparsity of the network by selecting these users. 

\begin{itemize}
    \item[\textit{Step 1.}] We initially crawled data from a small set of players to avoid exponential growth of the network. The Riot APIs allow collecting players in the leaderboard, having specified the tier and division of interest. Specifically, we retrieved the first page of the leaderboard for each of the four divisions of each tier (Bronze, Silver, and Gold). From those 2,460 users, we randomly selected a seed set of 100 players to avoid having a sparse network but rather a focused view on a part of the LoL population. For each player, we collected the information relative to the matches they played in one year, from September 4, 2019 - 2020. 
    
    \item[\textit{Step 2.}] We obtained the two team compositions from each match. Using this information, we built the player network by connecting the players participating in a match as teammates. This step led to a population of 81,954 players for EUW, 82,937 players for NA, and 59,663 players for KR. 
    \item[\textit{Step 3.}]  Before retrieving the 1-year matches information for the new players, we selected the Largest Connected Component (LCC) of each network, leading to 39396 nodes for EUW, 46421 nodes for NA, and 26550 nodes for KR. This filtering was required (a) to avoid an excessively sparse network and (b) due to the APIs' limitations. Retrieving the match information for each new player would have required an estimate of several years.
    
    \item[\textit{Step 4.}] We collected the match information for each new player and updated the network with the missing links, leading to 101,283 edges for EUW, 122,423 edges for NA, and 69296 edges for KR.
\end{itemize}

Table \ref{table:network_stat} summarizes the final networks' statistics describing the \#nodes, \#edges, density, average degree, and clustering for each region.

\begin{table}[]
\caption{Networks Statistics}
\begin{tabular}{lrrr}
\hline
                & \multicolumn{1}{l}{EUW} & \multicolumn{1}{l}{NA} & \multicolumn{1}{l}{KR} \\ \hline
\#nodes         & 39,396                  & 46,421                 & 26,550                 \\
\#edges         & 251,003                 & 626,451                & 111,795                \\
Density         & 0.0003                  & 0.0005                 & 0.0003                 \\
Avg. Degree     & 163.10                  & 146.86                 & 150.25                 \\
Avg. Clustering & 0.42                    & 0.19                   & 0.60                   \\ \hline
\end{tabular}
\label{table:network_stat}
\end{table}

As a result of the data collection phase, we obtained three networks representing each a geographical region. Each network models the implicit in-game interactions of the players: the players are connected if they have played as teammates in at least one match. The network is undirected, weighted, and dynamic. The weight represents how many matches two nodes played together as teammates. The dynamicity of the network derives from the matches having a timestamp. Hence, we can collocate a match, thus a link, at a specific point in time, throughout the  analyzed time window. A dynamic network allows evaluating the network's metrics evolution over time in a fine-grained way~\cite{Loria2020}. The continuity of dynamic networks can be discretized by representing the network as a sequence of several snapshots, without loss of information~\cite{Scripps2009}. Therefore, we narrowed our view to a week-level dividing the observation period into 52 snapshots. This division in weeks provides a general overview, comparable to the Covid-19 situation in each geographical area.

Having the network divided in weekly snapshots helps achieve a better understanding of how players' activities and properties evolved through time. We needed to bring those fine-grained data on a wider scale in order to compare them with Covid-19 evolution in various countries. The phases represented this further coarse-grained view. The phases group together different snapshots, which provided a way of understanding which period, during the year, needed to be compared based on the virus evolution.

\subsubsection*{Constructs and Metrics}
In this study, we collected and analyzed the telemetry data of LoL players, describing their activity and social bonds. \textit{The activity level of players} was computed in terms of the number of matches (\#matches) they played in the snapshot analyzed. We also measured how many players from our sample population were active (\#players). \textit{Player sociality} was described by metrics both at a player and network level. First, we evaluated the tendency of players to either play regularly with the same players or to play frequently with different players. We used the Focused Player metric, as defined in~\cite{Pirker2018} for the purpose. This value ranks players based on how they interact with others. Higher scores describe players who usually play with the same group of other players. Conversely, lower values are associated with players playing with different team members. The coefficient is calculated in the following way: 

\begin{center}
    \textit{FocusedPlayer} = $\frac{\#sum\_weights}{degree}\times\frac{\#match\_played}{\#tot\_match}$
\end{center}

The formula is also designed to eliminate biases that may be produced by highly active players. Besides this coefficient, we used network-based metrics, such as modularity, number of communities, and weighted clustering. Modularity describes the density of the connections in the network. When the modularity score is high, the network has dense connections within a community but sparse links among different communities. We computed the modularity score using the networxk~\footnote{\url{https://networkx.org/documentation/stable/reference/algorithms/generated/networkx.algorithms.community.quality.modularity.html}} python library. Finally, we measured the weighted clustering value. In SNA, the clustering coefficient measures the extent to which nodes tend to cluster together. Hence, how likely it is for nodes to form triangles. Weighted clustering embeds information on edges' weights in the formula.

We evaluated those metrics both at a network and an individual level. We analyzed values distribution at network level and compared how these values varied through the Covid-19 phases. For this purpose, we assessed a statistical difference between the network properties in the three lockdown phases using the Mann–Whitney U-test~\cite{Markus}. The hypothesis for this test is a 50\% probability that a randomly drawn member of the first population will exceed a member of the second population. The alternative null hypothesis can be double-tailed (the two samples come from the same population - i.e., both have the same median) or single-tailed (the values in one population are higher than the other). It is worth noting that we used the U-test since our data met its assumptions, namely: i) the dependent variable is measured at a continuous or ordinal level - e.g., number of matches, and the Focused Player score; ii) the independent variable consists of two categorical, independent groups, i.e., pre-lockdown vs. lockdown, and lockdown vs. post-lockdown; iii) the observations are independent, i.e., the groups are disjointed; and iv) the observations are not normally distributed. We fixed a p-value of $0.05$. At a player level, we tested whether the same features varied across the Covid-19 phases for each player. In other work, we measured how the featured players varied throughout the year. To this end we considered players who were retained in each phase, and gave them the name \textit{committed players}. Unlike the previous test (Mann-Withney-U), which compares the distribution of non-dependent samples, we used the Wilcoxon signed-rank test~\cite{Rey2011}, comparing dependent samples. In this case, the samples are matched. Hence data come from the same population. Usually, this test is used to evaluate whether a difference between pre and post-treatment exist. For instance, whether the metric's values varied among pre-lockdown and lockdown for each player.

Finally, we focused a magnifying glass on players entering our LoL network starting only from the lockdown phases: \textit{lockdown starters}. This category includes (i) new players, (ii) old players moved from another server, or (iii) players who did not play at least along the previous phase (pre-lockdown).

\section{Analysis and Results}
In this section, we outline the results of our analyses. First, we show how \textit{player turnout} changed across the different phases and networks. We then present the ways in which \textit{player activity levels} and \textit{sociality} changed. Finally, we describe some statistics and properties of \textit{lockdown starters}.

\subsection{Players' Turnout and Activity Level}
\begin{figure*}
\centering
\begin{subfigure}{0.50\linewidth}
\centering
\includegraphics[width=.8\linewidth]{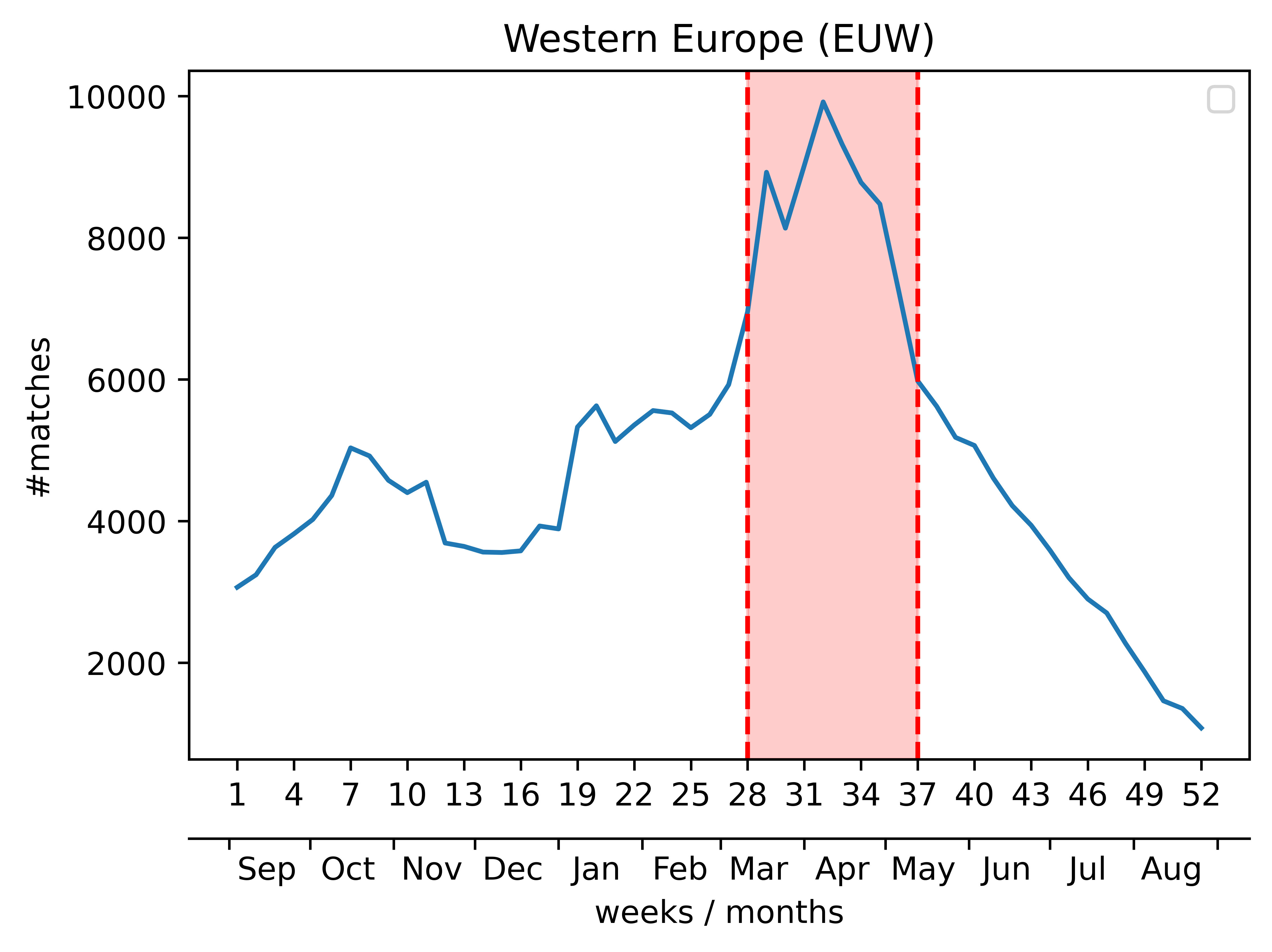}
\caption{}
\label{sfig:testa}
\end{subfigure}%
\begin{subfigure}{0.5\linewidth}
\centering
\includegraphics[width=.8\linewidth]{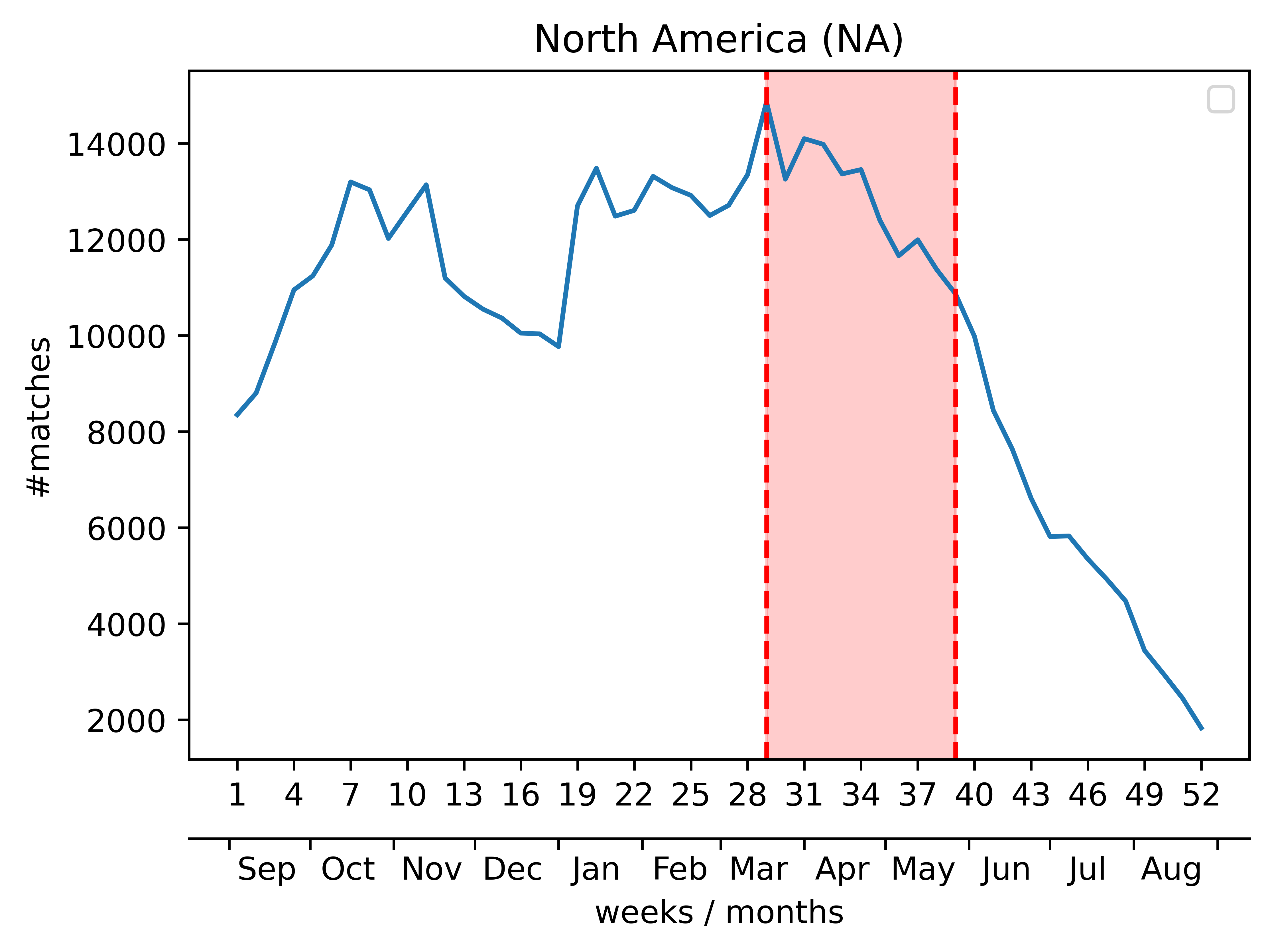}
\caption{}
\label{sfig:testb}
\end{subfigure}\par\medskip
\begin{subfigure}{0.5\linewidth}
\centering
\includegraphics[width=.8\linewidth]{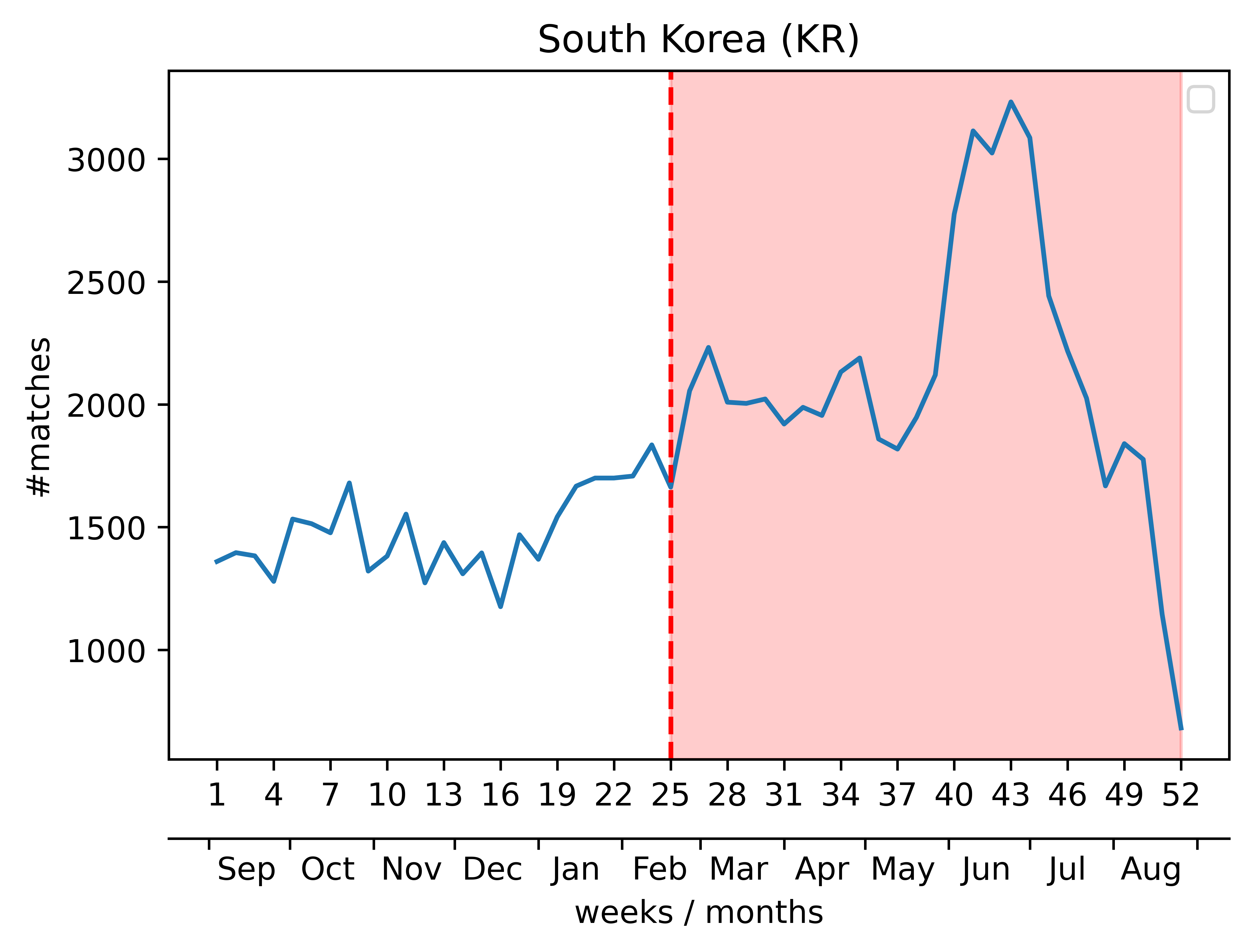}
\caption{}
\label{sfig:testc}
\end{subfigure}%
\caption{\textit{\#matches} across the whole year. The red area represents the lockdown phase. (a) EUW (b) NA (c) KR}
\label{fig:test}
\end{figure*}

Player turnout was modeled by measuring the number of matches and distinct-individual players in the weeks of each lockdown phase (i.e., pre-lockdown, lockdown, and post-lockdown). We conducted two sets of tests for each area. First, we built three (two in the case of Korea) populations: the number of matches in the pre-lockdown weeks (size = \#weeks in pre-lockdown), the number of matches in the lockdown weeks (size = \#weeks in lockdown), and the number of matches in the post-lockdown weeks (size = \#weeks in post-lockdown). Second, we built the same three (or in Korea two) populations considering the number of players. Figure~\ref{fig:test} shows the \textit{\#matches} trend in EUW, NA, and KR during the whole year. On approaching September (week 52), the number of matches quickly decreased for all the regions (week 52 matches: EUW 1078, NA 1827, KR 682).

\begin{table*}[]
\caption{EUW-NA-KR Matches distribution. Values are normalized to the number of weeks in each phase.}
\begin{tabular}{@{}llrrrrrrr@{}}
\toprule
             &                        & \textbf{Min} & \textbf{25\%} & \textbf{50\%} & \textbf{75\%} & \textbf{Max} & \textbf{Mean} & \textbf{Std} \\ \midrule
\textbf{}    & \textit{pre-lockdown}  & 0.037        & 0.037         & 0.111         & 0.296         & 30.66        & 0.369         & 1.024        \\
\textbf{EUW} & \textit{lockdown}      & 0.100        & 0.100         & 0.200         & 0.600         & 39.40        & 0.801         & 2.158        \\
\textbf{}    & \textit{post-lockdown} & 0.066        & 0.006         & 0.133         & 0.266         & 21.66        & 0.386         & 1.159        \\ \midrule
\textbf{}    & \textit{pre-lockdown}  & 0.035        & 0.107         & 0.285         & 0.714         & 22.85        & 0.683         & 1.253        \\
\textbf{NA}  & \textit{lockdown}      & 0.090        & 0.181         & 0.363         & 0.909         & 42.54        & 0.941         & 2.062        \\
\textbf{}    & \textit{post-lockdown} & 0.076        & 0.076         & 0.153         & 0.461         & 33.76        & 0.500         & 1.182        \\ \midrule
\textbf{}    & \textit{pre-lockdown}  & 0.041        & 0.041         & 0.083         & 0.208         & 17.04        & 0.274         & 0.807        \\
\textbf{KR}  & \textit{lockdown}      & 0.035        & 0.035         & 0.071         & 0.014         & 15.82        & 0.217         & 0.663        \\ \bottomrule
\end{tabular}
\label{table:match_dib}
\end{table*}

To assess a statistical differences among the different phases (or populations) in both \textit{\#matches} and \textit{\#players} we used the Mann–Whitney U-test~\cite{Markus}. Our wish was study wheter there is a significant difference between phases (\textit{p <.05}) by verifying the alternative single-tailed hypothesis (one population higher than the other).
The results show that the differences between the phases are statistical significant (\textit{p <.01}) concerning the \textit{\#matches} and \textit{\#players}. 

In Western Europe, there is a significant increment the \textit{\#matches} during the lockdown compared to the pre and post-lockdown phases (pre-lockdown and lockdown $U=0$ $p<.001$, lockdown and post-lockdown $U=150$ $p<.001$). In other terms, the \#matches in lockdown are significantly higher than both the \#matches in pre and post-lockdown. We also found a drop during post-lockdown compared to the pre-lockdown phase ($U=298$ $p<.001$). An analogous situation occurs in North America (pre-lockdown and lockdown $U=85$ $p<.001$, lockdown and post-lockdown $U=143$ $p<.001$, pre-lockdown and post-lockdown $U=359$ $p<.001$). In Korea, we also witnessed a rise in activity during lockdown compared to the pre-lockdown weeks ($U=61$, $p<.001$), but we lack information on the post-lockdown phase (see the division in phases, Section~\ref{subsec:covid_phases}).

Similar to the number of matches situation, Western Europe also saw an increase in the \textit{\#players}. The highest value for the number of individually distinct players was detected during the lockdown weeks, with a significant drop in the post-lockdown phase compared to the pre-lockdown weeks (pre-lockdown and lockdown $U=1$ $p<.001$, lockdown and post-lockdown $U=150$ $p<.001$, pre-lockdown and post-lockdown $U=300$ $p<.01$). Although the biggest turnout of individually distinct players in NA also occurred during the lockdown, the number of players in the pre-lockdown and post-lockdown weeks was not statistically different(lockdown and post-lockdown $U=143$ $p<.001$, pre-lockdown and post-lockdown $U=356$ $p<.01$). Finally, Korea experienced a peak of distinct players when restrictions were applied ($U=28$ $p<.01$).

Player activity levels describe how the number of matches changes across the Covid-19 phases for each player. Differently from players' turnout measuring players' activity at a global level, in these tests, we evaluate players' shifts in their activity. Towards this, we analyzed \textit{committed players}. Hence, these analyses show whether the playing habits of recurrent users were affected by the Covid-19 pandemic across the world. Compared to the whole population \textit{committed players} are: KR 8,593 (33.69\%), EUW 8,462 (21.47\%), NA 16,574 (35.70\%). 

Similar to the previous setting, we run a set of tests for each region. In this case, the three (or two) populations are paired. Hence, for the same region, the populations' sizes are the same and are equal to the number of \textit{committed players} within that area. Every player $p$ has a value in each population (pre-lockdown, lockdown, and post-lockdown). To assess a statistical difference among the different phases in the \#matches played by the \textit{committed players} we used the Wilcoxon signed-rank test~\cite{Rey2011}.  We wanted to verify: i) if there is a  significant difference between phases (\textit{p <.05}), and ii) verify the alternative single-tailed hypothesis (one population higher than the other). Table~\ref{table:match_dib} provides a summary of the match distribution in the different phases for each country. Values are normalized to the number of weeks in each phase (Section.~\ref{subsec:covid_phases}). Values are consistent with the previous findings. EUW had the greatest growth proportionally in player's activities (mean: pre-lockdown 0.369, lockdown 0.801) followed by NA. While in KR, the values suggest that players who were not playing many matches in pre-lockdown, on average, increase their activity more than the rest of the population (std: pre-lockdown 0.807, lockdown 0.663).

In Western Europe, we found that \textit{committed players} participated in a significantly greater number of matches during the lockdown (pre-lockdown and lockdown $U=0$ $p<.001$, lockdown and post-lockdown $U=150$ $p<.01$) . Their activity suffered a significant drop in the post-lockdown weeks. These results are coherent with the outcomes of the player turnout tests. An analogous situation occurs in North America(pre-lockdown and lockdown $U=85$ $p<.001$, lockdown and post-lockdown $U=143$ $p<.001$, pre-lockdown and post-lockdown $U=359$ $p<.01$) and in Korea (pre-lockdown and lockdown $U=61$ $p<.001$).

\begin{figure*}[h!]
\centering
\captionsetup{justification=centering}
\begin{subfigure}{0.50\linewidth}
\centering
\includegraphics[width=1\linewidth]{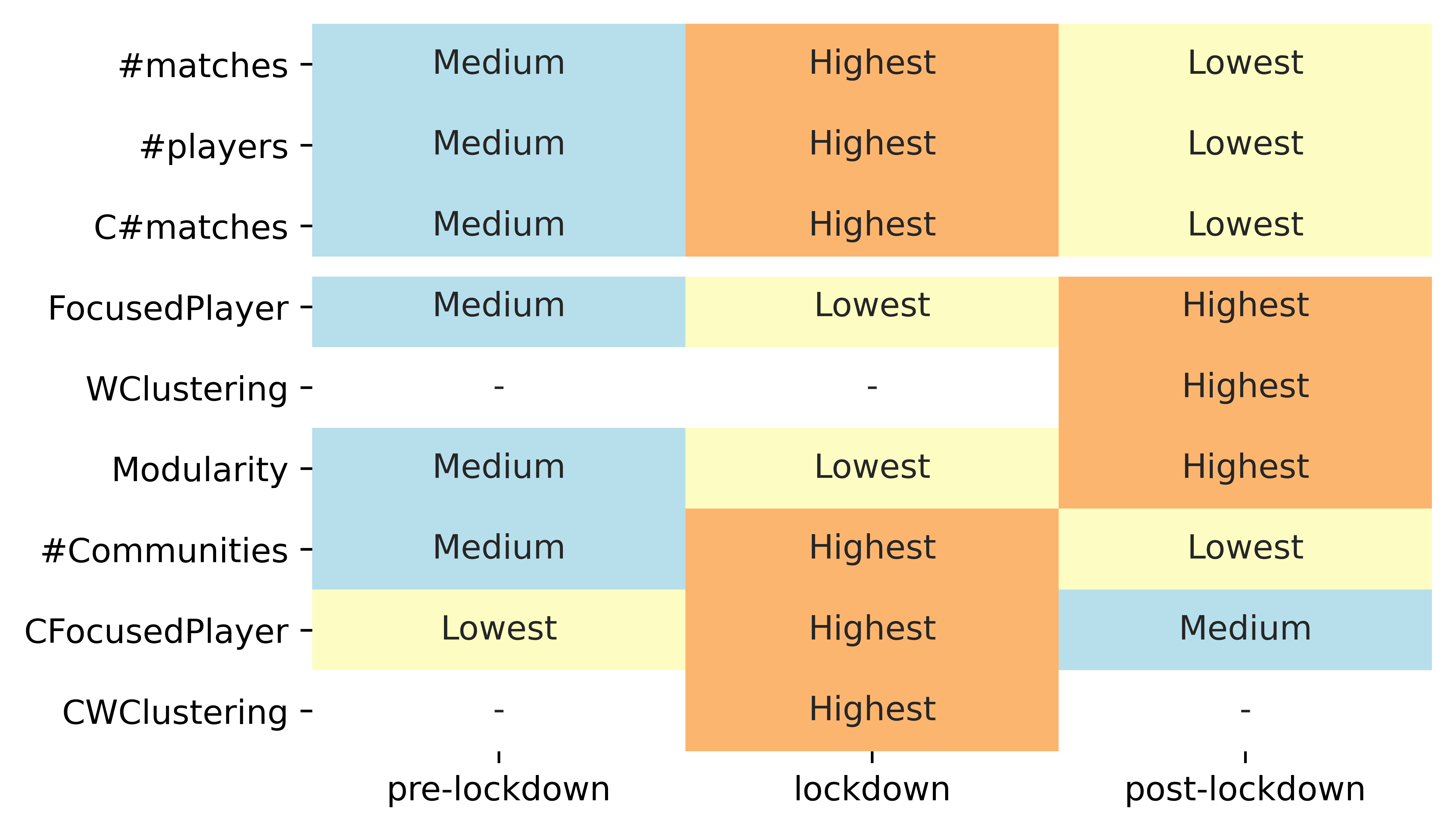}
\caption{}
\label{sfig:testa}
\end{subfigure}%
\begin{subfigure}{0.50\linewidth}
\centering
\includegraphics[width=1\linewidth]{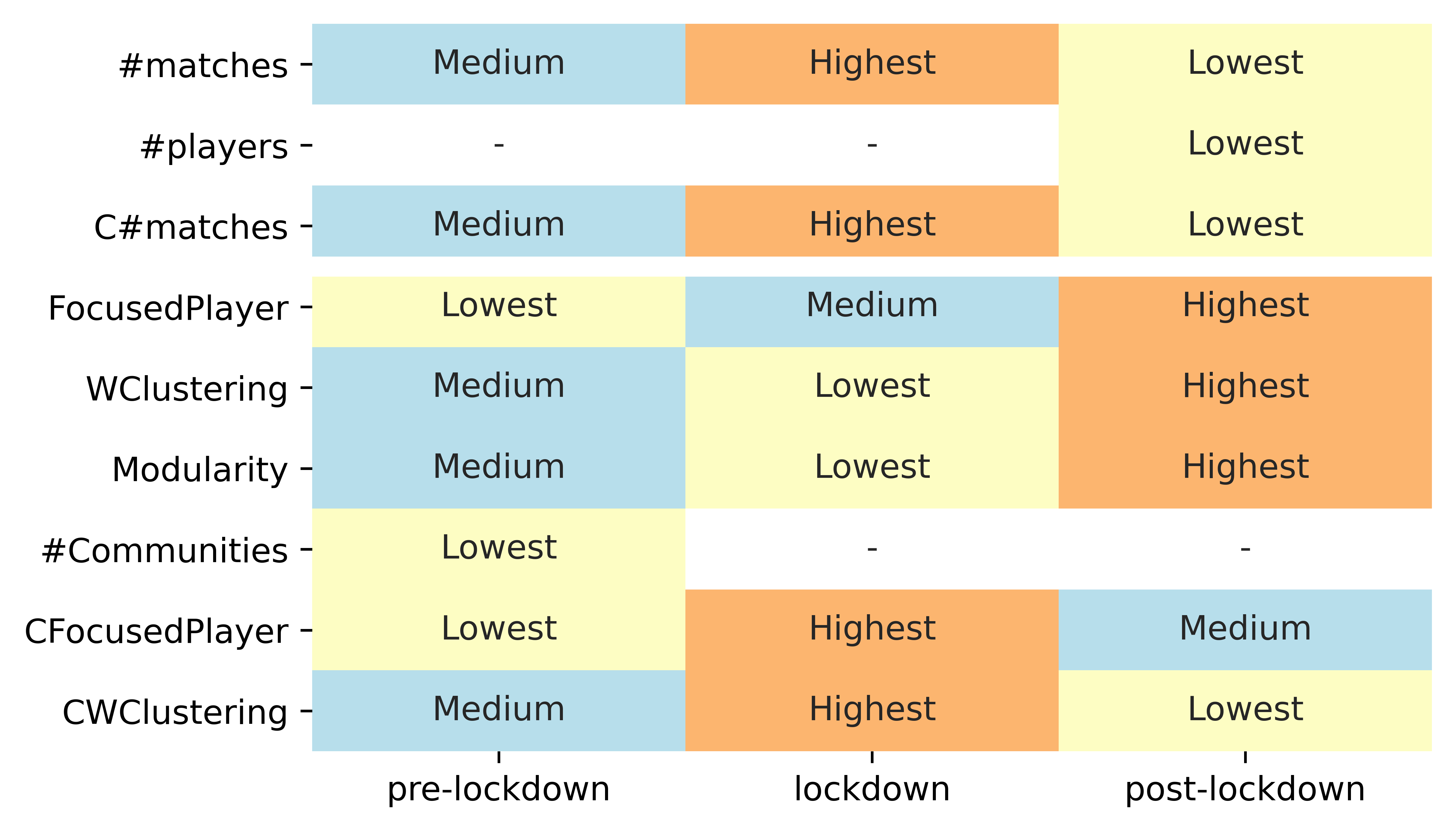}
\caption{}
\label{sfig:testb}
\end{subfigure}\par\medskip
\begin{subfigure}{0.50\linewidth}
\centering
\includegraphics[width=1\linewidth]{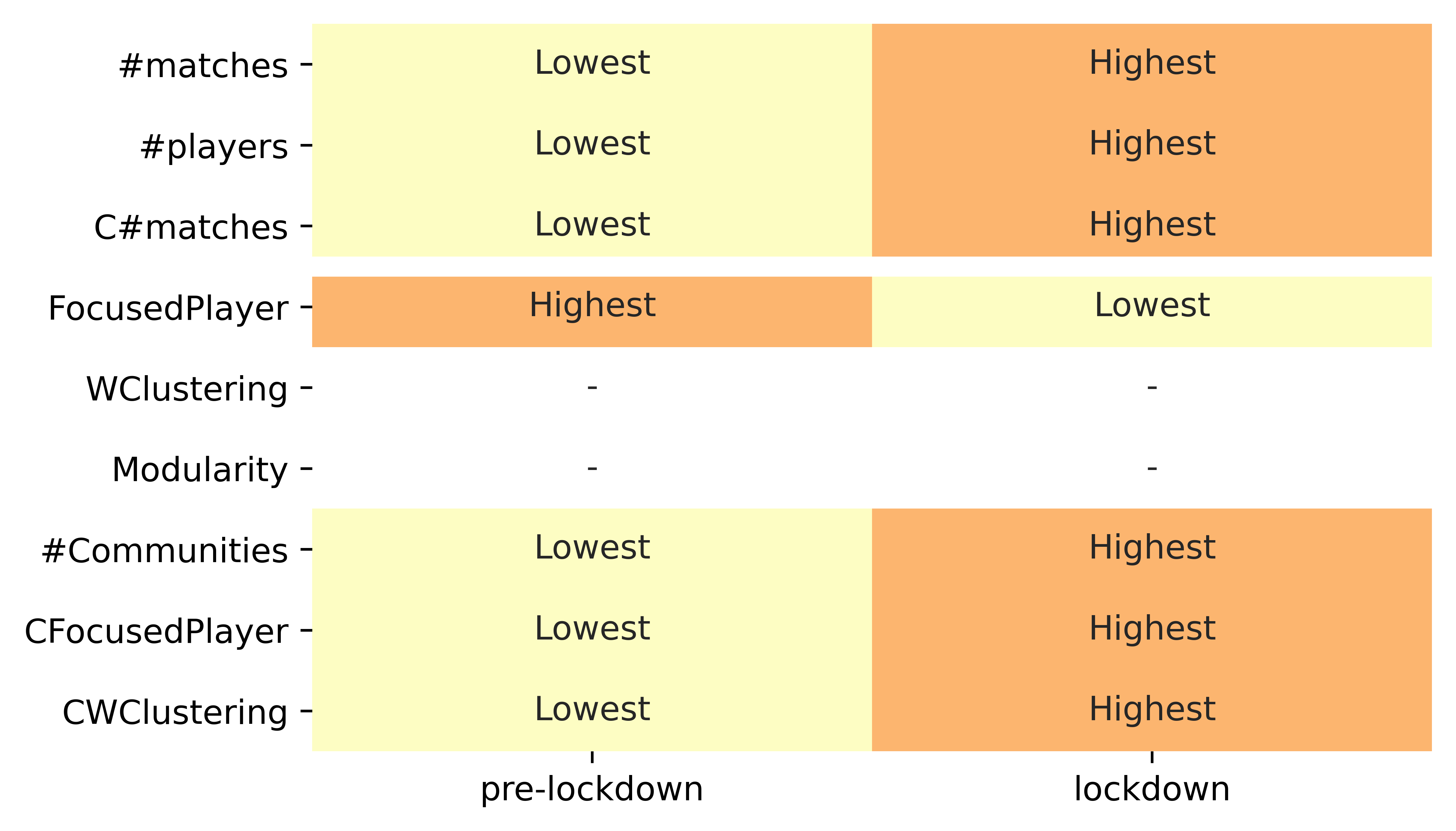}
\caption{}
\label{sfig:testc}
\end{subfigure}%
\caption{ Metrics evolution throughout the Covid-19 phases in (a) EUW (b) NA, and (c) KR. Te first three rows represent \textit{activity} metrics, while the remaining represents \textit{social} metrics. Note: i) cells with \textbf{-} indicate a non-significant difference between the two phases ii) metrics with \textbf{C} suffix are related to \textit{committed players}. For accessibility reasons, the color palette is color-blind friendly.}
\label{fig:metrics_heatmap}
\end{figure*}

Results are summarized in Figure~\ref{fig:metrics_heatmap}.

\subsection{Players' Sociality}
Player sociality is modeled using four different metrics: i) the FocusedPlayer value, ii) weighted clustering, iii) modularity, and iv) the number of different communities. We followed the same process as described in the previous analyses for player activity. First, we analyzed how sociality changed in the network using the  Mann–Whitney U-test~\cite{Markus}. We then verified the evolution of the same features across the different phases for \textit{committed players} using the Wilcoxon signed-rank test~\cite{Rey2011}. All the tests were repeated for each region (i.e., EUW, NA, and KR).

\par \noindent \textbf{Network-level Sociality.}
In both Western Europe and North America, the networks' \textit{FocusedPlayer} values grew in the post-lockdown weeks compared to the previous phases (pre-lockdown and post-lockdown). However, the two regions differ in that, in EUW, the lowest value was seen in the lockdown phase, whereas, in NA, the lowest scores were in the pre-lockdown weeks. In constrast to the other areas, Korea saw its highest FocusedPlayer values in the pre-lockdown. We obtained similar outcomes for the weighted clustering coefficient and the modularity score, which significantly increased in the post-lockdown phase for EUW and NA.  In Korea, the differences in weighted clustering and modularity are not statistically significant. Finally, we can see an increment in the \textit{\#communities} during the lockdown weeks in Europe and Korea. In North America, the number of communities dropped in the pre-lockdown weeks, like in Korea, while, in Europe, the lowest value was experienced in the post-lockdown phase. Results are summarized in Figure~\ref{fig:metrics_heatmap}.

\par \noindent \textbf{Committed Player Sociality.}
The results show that in all areas \textit{committed players} played more with the same people (higher FocusedPlayer values) than the rest of the network during the lockdown weeks than in the pre-lockdown phase. \textit{committed players} tend to play even more with the same players in the post-lockdown phase, leading to the highest FocusedPlayer values. This outcome can be observed both in the EUW (\textit{committed players} lockdown and post-lockdown $W=18,750,694$ $p<.001$, \textit{committed players} pre-lockdown and post-lockdown $W=17,323,162$ $p<.01$) and NA regions (\textit{committed players} lockdown and post-lockdown $W=74,610,718$ $p<.001$, \textit{committed players} pre-lockdown and post-lockdown $W=64,768,086$ $p<.001$), as we do not have a post-lockdown phase for Korea. In all regions, the weighted clustering value is at its highest during the lockdown weeks (EUW: \textit{committed players} pre-lockdown and lockdown $W=10,645,145$ $p<.001$, \textit{committed players} lockdown and post-lockdown $W=18,271,390$ $p<.001$ NA: \textit{committed players} pre-lockdown and lockdown $W=64,231,460$ $p<.001$, \textit{committed players} lockdown and post-lockdown $W=61,645,104$ $p<.001$.) There is no significant difference between pre and post-lockdown values for Western Europe, whereas North America experienced higher pre-lockdown values than in the post-lockdown weeks ($W=67,486,190$ $p<.01$). Finally, we could not compute modularity and number of communities for \textit{committed players}, as they are network-level measures, and local values for each player thus do not exist.

Results are summarized in Figure~\ref{fig:metrics_heatmap}.

\subsection{Lockdown Starters}~\label{subsec:lockdown_starters}
As introduced previously, \textit{lockdown starters} are players who started to play during the lockdown phase. In the following section, we show how substantial the presence of lockdown starters was in the Western Europe and North America populations. The total number of \textit{lockdown starters} is 7605 (19.35\%) for EUW and 4380 (9.45\%) for NA. We first studied the dropout rate, that is, the percentage of players who started to play during lockdown but did not continue to play in post-lockdown. The dropout rate is 54.43\% for EUW (remaining players 3465), and for NA, the rate is 35.02\% (remaining players 2846). 

We thendeepen our understanding of these players by investigating how they connect with other players. For this purpose, we compared the FocusedPlayer values for those \textit{lockdown starters} against the other players. We chose this metric because it represents the tendency of these players to play with the same players (i.e., strong, recurrent social bonds) versus the habit to play with different players (i.e., shallow social connections). We thus studied this value for \textit{lockdown starters}, also investigating \textit{lockdown starters} retained in the post-lockdown phase. Due to the lack of a post-lockdown phase in Korea, these analyses were only performed for the EUW and NA areas. We used the Mann-Whitney U test\cite{Markus} to compare the \textit{FocusedPlayer} values of \textit{lockdown starters} in the lockdown phase with: i) the rest of the network, ii) \textit{committed players} and iii) \textit{lockdown starters} retained in the post-lockdown phase. Then, we compared the \textit{FocusedPlayer} scores of retained \textit{lockdown starters} in the post-lockdown phase with i) the rest of the network and ii) \textit{committed players}.

\textbf{Lockdown starters}, in the lockdown weeks, played with different people (low FocusedPlayer scores) more than (i) the remainder of the network (EUW: $U=19,168,761$ $p<.001$, NA: $U=16,629,593$ $p<.001$), (ii) \textit{committed players} (EUW: $U=6,261,885$ $p<.001$, NA: $U=8,092,504$ $p<.001$), and (iii) \textit{lockdown starters} retained in the post-lockdown phase (EUW: $U=4,034,987$ $p<.001$, NA: $U=1,581,452$ $p<.001$), in both the EUW and NA networks.  Hence, \textit{lockdown starters} retained in the lockdown weeks tended to play more with the same people than all the \textit{lockdown starters}.

\textbf{Lockdown starters} retained in post-lockdown, for North America, played more with the same players (high FocusedPlayer scores) than the remainder of the network($U=33,284,423$ $p<.01$). Whereas, the same test was no statistically significant for Western Europe. Nevertheless, in both regions, \textit{committed players} played more together than retained \textit{lockdown starters} (EUW: $U=64,771,120$ $p<.001$, NA: $U=1,581,452$ $p<.01$).

\section{Discussion}
We identified changes in the LoL player activities and their social structure using SNA techniques. In the following, we discuss our findings in relation to the research questions we previously proposed (Section.~\ref{research_question}). Before talking about the limitations and future development of this work.

\subsection*{RQ1: How did player turnout and activity vary across the Covid-19 phases?}
The data shows a significant increment in the number of LoL matches and players during the lockdown weeks or movement restrictions. The same activity increase was found in \textit{committed players}' gaming behaviors. This finding is in alignment with the conception of games as a coping mechanism for social isolation~\cite{csener2021impact}, although it may also be related to a greater opportunity for playing due to the forced stay-at-home orders. While all geographical areas saw a peak in the number of matches, North America was the only area in which there was no significant increment in the number of players. Hence, in this area, the higher value for the number of matches during lockdown was almost solely due to the increase in activity of \textit{committed players}.

After the restrictions were lifted, in EUW and NA, the number of matches and players dropped significantly, also for \textit{committed players}. This drop provides a good description of the situation in those weeks that also corresponded with the start of the summer season. At that time people were eager to socialize in person and maybe even if possible to go on holiday. In this phase (post-lockdown), almost half of the players who joined in the lockdown weeks were lost (Section.~\ref{subsec:lockdown_starters}), accounting for between 10\% and 20\% of the whole population.

While we expected an activity peak and drop in the lockdown and post-lockdown phases, respectively, those results alone do not provide additional information on how LoL impacted social connections. Towards this, we studied the structure of the LoL implicit social network built from in-game interactions (i.e., participation in a match as teammates).

\subsection*{RQ2: How did player social structures vary across the Covid-19 phases?}

The huge waves of players in the lockdown phase, for Europe and Korea, generated a perturbation in the player networks. Both regions experienced significantly lower FocusedPlayer scores in the lockdown weeks. Hence, on average, the players population tended to play with more different users. This result should not be interpreted as a lack of player interest in forming recurrent bonds. It should rather be associated with the considerable increase in the number of players. As a consequence and from a global perspective, the matchmaking algorithm could choose among more players, leading to a greater diversity of teammates for players relying on automatic matching. This interpretation is supported by the analyses conducted on the North American network, where a peak in new players of this kind was not experienced during the lockdown. Here, we see an increase in the FocusedPlayer value compared to the pre-lockdown weeks. We can further support this thesis by investigating the FocusedPlayer scores for the \textit{committed players}. In all regions, the values were at their highest during the lockdown weeks for those players. This data confirms previous theories about games being incubators of social relationships~\cite{Pirker2018,Freeman2016MakingCommunity} and might be a sign that players fulfilled their need for belongingness~\cite{przybylski2010motivational} and connectedness~\cite{rogers2017motivational} in games. While some promising findings had already emerged from in the study of Animal Crossing Players, showing how the game fostered well-being and helped to cope with loneliness~\cite{johannes2020video,Mandryk2020HowWellbeing}. Nevertheless, LoL greatly differs from Animal Crossing in genre and purpose. Our study provides initial evidence that games which do not focus primarily on social mechanics and that mimic real-world experience can also connect players socially. 
 
The analysis of social bonds in the post-lockdown phase (Western Europe and North America) also brought interesting outcomes. Players' showed the highest values in the FocusedPlayer, Weighted Clustering, and Modularity metrics. As a result some players not only strengthened their connections during the lockdown, but their feelings of connectedness remained even after the restrictions were lifted. We have to acknowledge that we also saw a  player population decline in post-lockdown, which might have impacted the results for an argument inverse to the one previously presented. The automatic matching algorithm has fewer players to choose from. Yet, \textit{committed players} had a higher FocusedPlayer value in the post-lockdown phase than they had in pre-lockdown. In other words, committed players found a social platform in games when they missed real-world social contacts (lockdown), this feeling decreased slightly once they had the opportunity to meet in reality (post-lockdown) but it was still strong than it had been previously (pre-lockdown).

\subsection*{RQ3: Who are the players that approached the game during the Covid-19 lockdowns ?}
Our in-depth analysis on players joining the network during the lockdown (i.e., \textit{lockdown starters}), we provided further support for previous research showing how sociality affects retention~\cite{Pirker2018}. \textit{Lockdown starters} had tendentially lower values of FocusedPlayer than the other players in the network, probably because they entered the game later. Nevertheless, \textit{lockdown starters} retained in the post-lockdown weeks (roughly 50\%) showed significantly higher FocusedPlayer scores than those who dropped out. As a result players who found more stable connections (teammates) were more likely to be retained even when the restrictions were lifted. In North America, we cam see an even stronger result in this respect. \textit{Lockdown starters} retained in the post-lockdown weeks showed a stronger FocusedPlayer score than other players retained in the lockdown phase. This result is probably due to the increase in the number of players during lockdown (\textit{lockdown starters}) being less sensitive than in Europe--i.e., not statistically significant. Therefore, we can conclude that \textit{lockdown starters} who were retained were indeed involved in more stable connections than those who dropped out, but they were not necessarily involved in stronger bonds than other retained players (joined before the lockdown). 
 
\subsection*{RQ4: How have player networks varied across different geographical regions?}
Despite having already compared the findings across the different geographical areas, we will provide more in-depth observations on their differences and similarities in the following. The test results from the three areas overlapped in several situations. For instance, we saw a significant increment in player activity during lockdowns (and other restrictions) everywhere. Nevertheless, the same player peak seen in EUW and KR does not occur in North America. A peak of this kind is readily explained in Western Europe, where the restrictions were stringent, and in some countries even more so than in others (e.g., France, Italy, and Spain). Although the restrictions were less stringent in Korea, we observed a similar increase, which might be imputed not to the fear of the virus but to a feeling of responsibility linked to the transparency with which the pandemic was handled by the Korean government~\cite{Lee2020}. Furthermore, unlike other regions in KR we can see a sudden peak of matches between week 40-46. This rapid growth does not appear to be related to an increase in restrictions or cases but rather to a major LoL online event \cite{esguerra_2020}. Finally, in North America, the increase in the number of players was not as significant, despite the restrictions that existed. This finding should also be contextualized with the political and social conditions. In the U.S, which represents the majority of NA area, the beginning of the pandemic was taken less seriously due to some political statements \cite{vinopal_2020}. Moreover, the differences in the U.S states in terms of restriction strictness might have influenced the population in different ways. Whereas in issues concerning sociality, all the regions witnessed the desire on the part of a portion of players to connect with others and build recurrent bonds in the network, which would be durable in time (post-lockdown). 

\subsection{Limitations}
Our work has a number of limitations. First, the study is based on a single game. Second, our results are strongly related to the phase division. We can argue that a different phase division could lead to different results. In the EUW area, the restrictions are well documented and defined within a specific period. The information from NA and KR is less accurate. To mitigate this problem, we begin/end each phase only when the majority of countries/states/provinces in that area started/ended most of the restrictions (related to shops, restaurants, sports, and entertainment activities) or personal limitations (like gathering/traveling bans and stay-at-home orders). Moreover, potential events external to the pandemic, which may have altered the data in some weeks, were hard to detect, especially for South Korea. Third, we could not define a precise post-lockdown phase for KR due to the different approach to the Covid-19 outbreak taken by South Korea's government. Fourth, our networks were developed from small seed sets of players, although these represented active and committed players.

\subsection{Future Works}
While this work takes a further step forwards for the understanding of game player social behaviors, there is still much room for future studies. For instance, further and detailed research can be done on the evolution of the communities and their composition during the different phases, and which were the characteristics of the most cohesive groups. We can also expand our findings by conducting a more comprehensive behavioral analysis to detect the properties and characteristics of \textit{committed players} and \textit{lockdown starters}, focusing on players' forming strong, recurrent bonds. Moreover, while, in this study, we followed a high-level approach, analyzing statistical differences among groups of the population in the Covid-19 stages, we can run player-centric investigations analyzing the evolution of social behaviors. Finally, we can extend the observation period by analyzing and comparing the Covid-19 \textit{second wave} to the \textit{first wave}.

\section{Conclusions}
Social distancing and remote working (and studying) drastically changed people's lives, disrupting their routines and basic social interactions. The \#StayAtHome mantra, which recurred throughout the past year, brought additional impetus and force to the researching of tools that both entertain and connect. Games, for example, saw unprecedented growth and in some cases these were also proven to reduce anxiety and produce a sense of well-being in their players. This rise is coherent with recent research showing how multiplayer games foster connectedness and enable real-world friendships. In this study, we expand the analysis focusing on how social interaction patterns and the structure of the player social networks were affected by Covid-19 lockdowns. We modeled three different League of Legends player networks, each representing a different geographical area throughout a whole year (September 2019-2020). Our findings confirmed this activity and player rise already detected in other game genres, as well as the formation of stronger social connections for some players. \textit{Committed players} formed more recurrent and narrower bonds during the lockdown weeks. Additionally, new players forming stronger connections during lockdowns were also more likely to be retained in the game when the restrictions were lifted. These findings were similar across the three geographical areas, where small specific differences might also result from political and cultural differences. In conclusion, our findings support the idea that online games can help tackle boredom, loneliness and foster social interactions, even more so in times when real-world connections are limited.

\section{Acknowledgment}
This research is supported by funds of the state of Styria, Science and Research (Land Steiermark, Wissenschaft und Forschung).
\bibliographystyle{ACM-Reference-Format}
\bibliography{_main}

\end{document}